\documentclass[aps,prl,twocolumn,superscriptaddress,showpacs]{revtex4}
\usepackage{graphicx}

\usepackage{dcolumn}

\usepackage{bm}
\usepackage{color}
\usepackage[colorlinks]{hyperref}
\input{epsf}

\begin{document}


\title{Radiation induced oscillatory Hall effect in high mobility GaAs/AlGaAs
devices}

\author{R. G. Mani}
\author{V. Narayanamurti}
\affiliation {Harvard University, Gordon McKay Laboratory of
Applied Science, 9 Oxford Street, Cambridge, MA 02138, USA}

\author{K. von Klitzing}
\author{J. H. Smet}
\affiliation {Max-Planck-Institut f\"{u}r Festk\"{o}rperforschung,
Heisenbergstrasse 1, 70569 Stuttgart, Germany}

\author{W. B. Johnson}
\affiliation {Laboratory for Physical Sciences, University of
Maryland, College Park, MD 20740, USA}

\author{V. Umansky}
\affiliation{Braun Center for Submicron Research, Weizmann
Institute, Rehovot 76100, Israel}
%
%
%
%
\date{\today}
\begin{abstract}
We examine the radiation induced modification of the Hall effect
in high mobility GaAs/AlGaAs devices that exhibit vanishing
resistance under microwave excitation. The modification in the
Hall effect upon irradiation is characterized by (a) a small
reduction in the slope of the Hall resistance curve with respect
to the dark value, (b) a periodic reduction in the magnitude of
the Hall resistance, $R_{xy}$, that correlates with an increase in
the diagonal resistance, $R_{xx}$, and (c) a Hall resistance
correction that disappears as the diagonal resistance vanishes.
\end{abstract}
%
%
\maketitle Vanishing electrical resistance has served to introduce
new physical phenomena in condensed matter physics such as, for
example, quantum Hall effects (QHE), which stemmed from studies of
zero-resistance states at low temperatures $(T)$ and high magnetic
fields $(B)$ in the 2-Dimensional Electron System
(2DES).\cite{1,2} Recently, the high mobility 2DES provided an
unexpected surprise by exhibiting novel zero-resistance states
upon irradiation by low energy photons. In this instance,
vanishing diagonal resistance occurred about $B$ = $(4/5) B_{f}$
and $B$ = $(4/9) B_{f}$, where $B_{f}$ = $2\pi f m^{*}/e$, $m^{*}$
is an effective mass, $e$ is the electron charge, and $f$ is the
radiation frequency, while the resistance-minima followed the
series $B$ = $[4/(4j+1)] B_{f}$ with $j$=$1,2,3$...\cite{3}
Remarkably, vanishing resistance induced by microwave excitation
of the 2DES did not produce plateaus in the Hall resistance,
although the diagonal resistance exhibited activated transport and
zero-resistance states, similar to QHE.\cite{3,4} These striking
features have motivated substantial theoretical interest in this
phenomenon.\cite{5,6,7,8,9,10}

It is well known from experimental studies of transport that
changes in the diagonal conductivity, $\sigma_{xx}$, such as those
induced by the radiation in this context, can produce small
corrections, in the strong field limit, in the Hall resistivity,
$\rho_{xy}$, via the tensor relation $\rho_{xy} =
\sigma_{xy}/(\sigma_{xx}^{2} +\sigma_{xy}^{2})$.\cite{11}. Durst
and coworkers have also mentioned the possibility of an
oscillatory Hall effect from their theoretical
perspective.\cite{7} Experimental results reported here examine in
detail a radiation-induced modification of the Hall effect, which
occurred in Fig. 1(b) of the article from the year 2002 of ref. 3,
and Fig. 1(b) of ref. 12. In particular, it is demonstrated that
microwave excitation changes the slope of the Hall resistance,
$R_{xy}$, vs. $B$ curve by $\leq$ $1.5\%$. Further, there appears
to be an oscillatory variation in $R_{xy}$, where a reduction, in
magnitude, of the Hall resistance correlates with an increase in
the diagonal resistance, $R_{xx}$. It is also demonstrated that
the correction to the Hall resistance disappears as $R_{xx}
\longrightarrow 0$. Finally, the oscillatory variation in the Hall
resistance is comparable, in magnitude, to the radiation induced
change in the diagonal resistance, although the change, $\Delta
R_{xy}$, is small ($\leq$ 5$\%$) in comparison to $R_{xy}$.

Measurements were performed on standard devices fabricated from
GaAs/AlGaAs heterostructure junctions. After a brief illumination
by a red LED, the best material was typically characterized by an
electron density, $n$(4.2 K) $\approx$ $3$ $\times$ $10^{11}$
cm$^{-2}$, and an electron mobility $\mu$(1.5 K) $\approx$ $1.5$
$\times$ $10^{7}$ cm$^{2}$/Vs. Lock-in based four-terminal
electrical measurements were carried out with the sample mounted
inside a waveguide and immersed in pumped liquid He-3 or He-4, as
the specimens were excited with electromagnetic (EM) waves in the
microwave part of the spectrum, $27 \leq f \leq 170$ GHz. In this
report, we illustrate the characteristics of the radiation induced
modification of the Hall effect and point out a similarity to the
situation in the quantum Hall limit.
\begin{figure}
\begin{center}\leavevmode
\includegraphics[scale = 0.25,angle=0,keepaspectratio=true,width=2.75in]{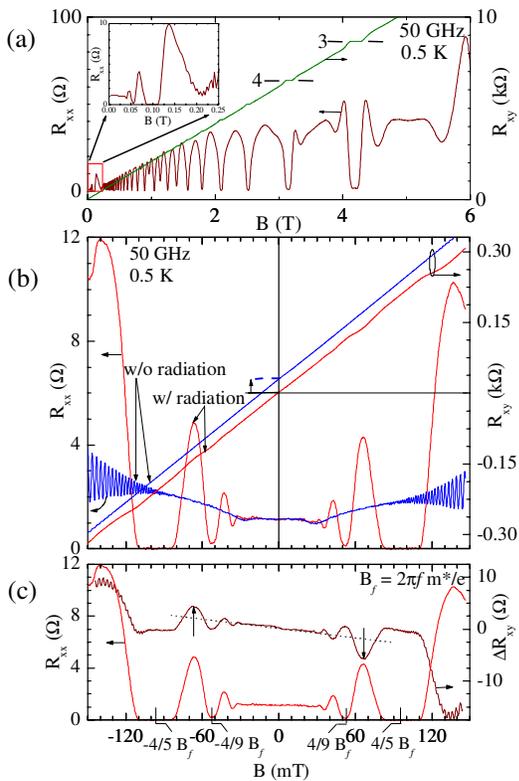}
\caption{(color online) (a): The Hall ($R_{xy}$) and diagonal
($R_{xx}$) resistances are plotted vs. the magnetic field, $B$,
for a GaAs/AlGaAs device under excitation at 50 GHz. Quantum Hall
effects (QHE) occur at high B as $R_{xx}$ vanishes. (inset): An
expanded view of the low-B data. (b): Data over low magnetic
fields obtained both with (w/) and without (w/o) microwaves at 50
GHz. Here, radiation induced vanishing resistance about $(4/5)
B_{f}$ does not produce plateaus in the Hall resistance, unlike in
QHE. Yet, an edgewise inspection reveals that there are
antisymmetric-in-B oscillations in $R_{xy}$ that correlate with
the $R_{xx}$ oscillations. The w/o radiation Hall data have been
offset here for the sake of clarity. (c): A comparison of the
radiation induced $R_{xx}$ oscillations with the radiation induced
change, $\Delta R_{xy}$, in the Hall resistance. Note the finite
slope (dotted line) in $\Delta R_{xy}$, and the odd symmetry under
field reversal, which is characteristic of the Hall effect. }
\label{figurename1}\end{center}\end{figure}

\begin{figure}
\begin{center}
\includegraphics*[scale = 0.25,angle=0,keepaspectratio=true,width=3in]{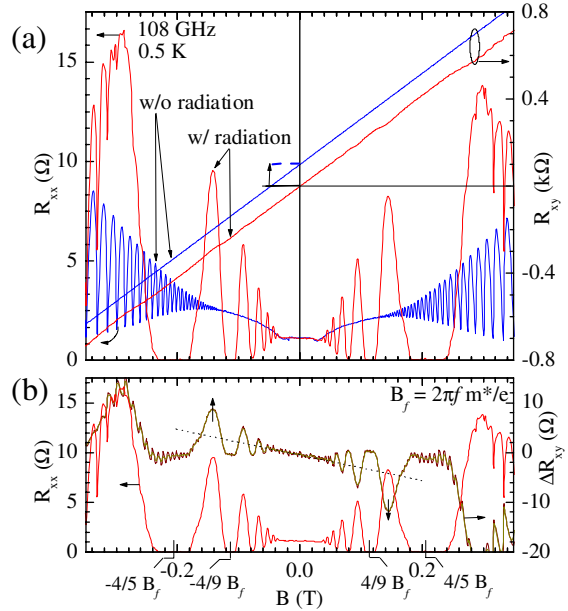}
\caption{(color online) (a): Data over low magnetic fields
obtained both with (w/) and without (w/o) microwave radiation at
108 GHz. Here, once again, radiation induced vanishing resistance
about $(4/5) B_{f}$ and $(4/9) B_{f}$ does not produce plateaus in
the Hall resistance, although antisymmetric-in-B oscillations in
$R_{xy}$ correlate with the $R_{xx}$ oscillations. The slope of
the $R_{xy}$ is reduced by the radiation. The w/o radiation Hall
data have been offset for the sake of clarity. (b): The radiation
induced portion of the Hall resistance, $\Delta R_{xy}$, is shown
along with $R_{xx}$. There is a slope to the $\Delta R_{xy}$ curve
(dotted line) because the radiation reduces the slope of the Hall
curve in (a). The correction to the Hall resistance, $\Delta
R_{xy}$, vanishes over the $B$-interval of the $R_{xx}$
zero-resistance states. Shubnikov-de Haas type resistance
oscillations, which are observable in the $\Delta R_{xy}$ trace in
the vicinity of the $(4/5) B_{f}$ zero-resistance state, are
attributed to the dark signal.}\label{1}
\end{center}
\end{figure}
Fig. 1 (a) shows measurements of the diagonal ($R_{xx}$) and Hall
$(R_{xy})$ resistances where, under microwave excitation at 50
GHz, $R_{xx}$ and $R_{xy}$ exhibit the usual quantum Hall behavior
for $B$ $\geq$ 0.3 Tesla.\cite{1,2} In contrast, for $B$ $<$ 0.25
Tesla, see inset of Fig. 1(a), a radiation induced signal occurs
and the resistance vanishes over a broad B-interval about $B$ =
0.1 Tesla. Further high-resolution measurements are shown in Fig.
1 (b). Without EM-wave excitation, $R_{xx}$ exhibits
Shubnikov-deHaas oscillations for $B$ $>$ 100 milliTesla (Fig. 1
(b)). The application of microwaves induces resistance
oscillations, which are characterized by the property that the
$R_{xx}$ under radiation falls below the $R_{xx}$ without
radiation, over broad $B$-intervals.\cite{3,4,12} Indeed, $R_{xx}$
appears to vanish about $(4/5) B_{f}$.\cite{3} Although these
zero-resistance-states exhibit a flat bottom as in the quantum
Hall regime,\cite{1,2} $R_{xy}$ under radiation does not exhibit
plateaus over the same $B$-interval.

Yet, the $R_{xy}$ data of Fig. 1(b) show perceptible oscillations
in the Hall resistance that are induced by the
radiation.\cite{3,12,13} Indeed, an edgewise inspection of the
data shows that there is an antisymmetric oscillatory component in
$R_{xy}$, in addition to a small radiation-induced change in the
slope of the Hall curve.\cite{12}  In order to highlight these
changes in the Hall effect, the radiation induced portion of the
Hall resistance, $\Delta R_{xy} = R_{xy}^{excited} -
R_{xy}^{dark}$, is shown along with $R_{xx}$ in Fig. 1(c). Here,
$R_{xy}^{dark}$ is the Hall resistance obtained without (w/o)
radiation, and $R_{xy}^{excited}$ is the Hall resistance with (w/)
radiation. As this procedure for extracting the radiation induced
Hall resistance involves the subtraction of two large signals,
$R_{xy}^{dark}$ and $R_{xy}^{excited}$, and since $\Delta R_{xy}$
is only a few percent of the dark Hall signal, it is necessary to
realize a stable low-noise experimental setup that minimizes
parameter drifts with time in order to obtain noise-free reliable
$\Delta R_{xy}$ data. The plot of Fig. 1(c) confirms a robust
$\Delta R_{xy}$ signal. Indeed, the observed characteristics in
Fig. 1(c), namely, a $\Delta R_{xy}$ signal that vanishes as $B
\longrightarrow 0$, and odd symmetry in $\Delta R_{xy}$ under
field reversal, help to rule out the possibility that $\Delta
R_{xy}$ originates from a mixture of the diagonal resistance with
$R_{xy}$, as a result of a misalignment of the Hall voltage
contacts. Further, the observation of quantized Hall effect in
Fig. 1(a) under radiation, and the
radiation-intensity-independence of the period of the radiation
induced oscillations, helps to rule out parallel conduction as the
origin of non-vanishing $\Delta R_{xy}$. These resistances scale
to resistivities with a scale factor of one.
\begin{figure}
\begin{center}
\includegraphics*[scale = 0.25,angle=0,keepaspectratio=true,width=2.75in]{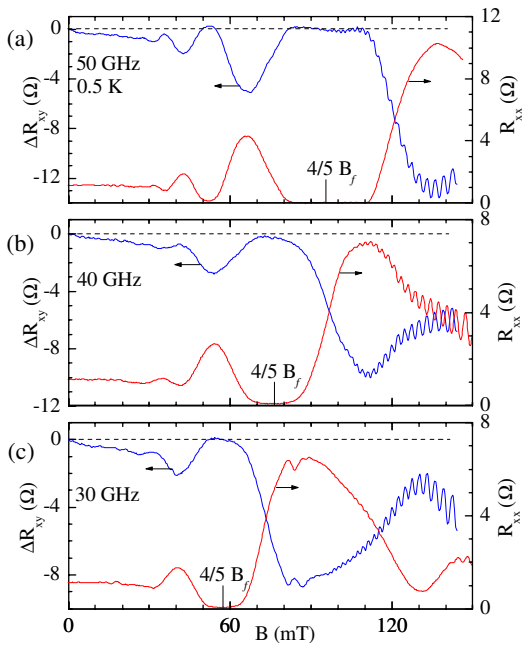}
\caption{(color online) The radiation induced change in the Hall
resistance, $\Delta R_{xy}$, is shown on the left ordinate and the
diagonal resistance, $R_{xx}$, is shown on the right ordinate for
microwave excitation at (a) 50 GHz, (b) 40 GHz, and (c) 30 GHz.
Here, $\Delta R_{xy}$ tends to vanish as the diagonal resistance
becomes exponentially small in the vicinity of $(4/5) B_{f}$. In
addition, a finite slope is evident in $\Delta R_{xy}$ for $B \leq
30$ mT, as in Figs. 1(c) and 2(b).}\label{1}
\end{center}
\end{figure}

The data of Fig. 2 illustrate this same effect when the specimen
is excited with microwaves at $f$ = 108 GHz. In Fig. 2(a), the
Hall data without radiation have been offset with respect to the
Hall data with radiation in order to bring out the oscillatory
portion of the Hall effect. Once again, an edgewise inspection of
the irradiated $R_{xy}$ data identifies an anti-symmetric
oscillatory Hall resistance, while it also reveals a small
reduction in the slope of the Hall resistance. A comparison of
Fig. 2(a) with Fig. 2(b) shows clearly that $R_{xy}$ is reduced in
magnitude over the B-intervals where $R_{xx}$ is enhanced by the
radiation. On the other hand, as the diagonal resistance vanishes
upon microwave excitation, as in the vicinity of $B = (4/5)
B_{f}$, for example, the correction $\Delta R_{xy}$ also vanishes.
The dotted line in the plot of $\Delta R_{xy}$ in Fig. 2(b) also
confirms that there is an approximately 1.5$\%$ reduction in the
slope of $R_{xy}$ that is induced by the radiation. The small
field reversal asymmetry that is observable in the w/ radiation
$R_{xx}$ data of Fig. 2 could be related to this radiation induced
change in the slope of $R_{xy}$. All these features are also
observable in Figs. 1(b) and 1(c). Our studies indicate that the
radiation induced reduction in the slope of the Hall curve brings
with it a corresponding small shift (at the per-cent level) in the
positions of the extrema in the Shubnikov-de Haas (SdH)
oscillations to higher magnetic fields. A straightforward
interpretation suggests that the change in the $R_{xy}$ slope, and
the shift of the SdH extrema, result from a radiation induced
change in the cross-sectional area of the Fermi surface.
\begin{figure}
\begin{center}
\includegraphics*[scale = 0.25,angle=0,keepaspectratio=true,width=2.75in]{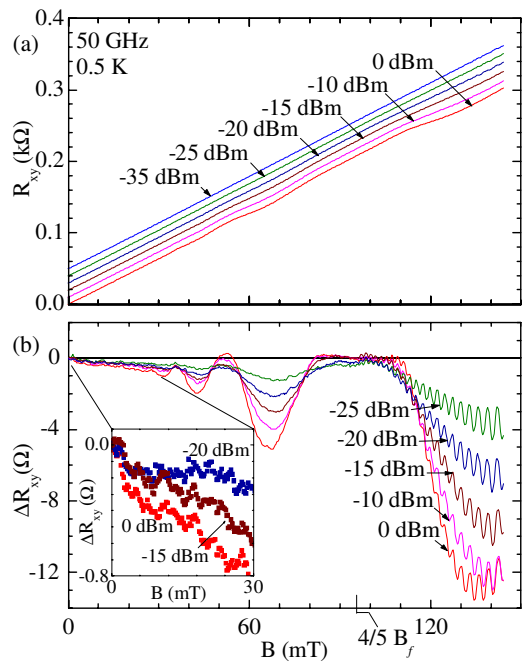}
\caption{(color online) (a): The Hall resistance $R_{xy}$ is shown
as a function of the magnetic field $B$ for radiation intensities,
given in units of dBm. The data indicate progressively stronger
modulations in $R_{xy}$ with increased radiation intensity. Data
have been offset for the sake of clarity. (b): This plot shows the
radiation-induced change in the Hall resistance $\Delta R_{xy}$
obtained from the data of (a). The inset illustrates the $B$
dependence of $\Delta R_{xy}$ for $B \leq 30$ mT.}\label{1}
\end{center}
\end{figure}

Figure 3 illustrates the frequency dependence, and the correlation
between $\Delta R_{xy}$ and $R_{xx}$, at microwave frequencies $f$
= 30, 40, and 50 GHz. Here, and also in Fig. 4, the exhibited Hall
resistances represent an antisymmetric combination of the signal
obtained for the two directions, ($+B, -B$), of the magnetic
field. In each case shown in Fig. 3, it is observed that (a) the
magnitude of $\Delta R_{xy}$ is approximately the same as the
magnitude of $R_{xx}$, (b) an increase in $R_{xx}$ under microwave
excitation leads to a decrease in the magnitude of $R_{xy}$, i.e.,
$\Delta R_{xy} \leq 0$ for $B \geq 0$, (c) $\Delta R_{xy}$ shows a
finite slope as $B \longrightarrow$ 0, and (d) the correction
$\Delta R_{xy}$ vanishes as the diagonal resistance vanishes in
the vicinity of $B = (4/5) B_{f}$. These data also confirm that
the characteristic field scale for the oscillatory Hall effect
follows the microwave frequency in the same way as $R_{xx}$.

We exhibit the power dependence of the Hall resistance at a fixed
microwave frequency in Figure 4. Fig. 4(a) shows $R_{xy}$ for
various microwave intensities in units of dBm, with 0 dBm = 1 mW.
Here, it is evident that increasing the radiation power produces
progressively stronger variations in $R_{xy}$. This is confirmed
by Fig. 4(b), which demonstrates that the minima in $\Delta
R_{xy}$ become deeper with increased power, consistent with the
results of Fig. 1(c), Fig. 2(b), and Fig. 3.

There appear to be some similarities between the Hall resistance
correction reported here and the experimental observations under
quantum Hall conditions. In the vicinity of integral filling
factors in the quantum Hall situation, the measured Hall
resistivity, $\rho_{xy}$, is expected to approach the quantum Hall
resistance, $R_{H}(i) = h/(e^{2}i)$, in the limit of vanishing
diagonal resistivity, i.e., $\rho_{xx} \longrightarrow 0$, at
zero-temperature.\cite{14} At finite temperature, experiment has
indicated corrections to the Hall resistivity that are
proportional to the magnitude of the diagonal resistivity, $\Delta
\rho_{xy} = - s\rho_{xx}$, with $s$ a device dependent
constant.\cite{14} That is, a finite (non-vanishing) $\rho_{xx}$
can lead to a reduction in the magnitude of $\rho_{xy}$. This is
analogous to the effect reported here since an increase in
$R_{xx}$ due to microwave excitation leads also to a decrease in
the magnitude of the Hall resistance, i.e., $\Delta R_{xy} \sim -
R_{xx}$ in Fig. 3. And, as the $R_{xx}$ vanishes, so does the
correction to the Hall resistance. Indeed, experiment suggests
some device dependence to this radiation induced Hall resistance
correction just as in the quantum Hall situation.\cite{14} Thus,
one might suggest that increased backscattering due to the
radiation,\cite{5,7,9} leads to an increased dissipative current
over the $R_{xx}$ peaks, which loads the Hall effect, similar to
suggestions in ref. 14.\cite{14} It might then follow that the
suppression of the backscattered current over the $R_{xx}$ minima,
eliminates also the correction to the Hall effect in our
experiment.

The radiation induced change in the slope of the Hall curve
appears to fall, however, outside the scope of this analogy, and
this helps to identify the dissimilarities between the
radiation-induced effect reported here and the quantum Hall
situation: In the quantum Hall situation, the effect of finite
temperature is to mainly increase the diagonal resistance, which
leads to a decrease in the observed Hall resistance.\cite{14} In
the case of the radiation induced resistance oscillations, the
radiation, which plays here a role that is similar to the
temperature in the quantum Hall situation, can serve to both
increase and decrease the diagonal resistance. Theory would
suggest the existence of both a downhill- and uphill- radiation
induced current with respect to the Hall field.\cite{7} Naively,
one might then expect both an enhancement and a diminishment of
the magnitude of the Hall resistance if the quantum Hall analogy
carried over to this case. Yet, we have found that the magnitude
of $R_{xy}$ is mainly reduced by the radiation. The radiation
induced change in the slope of the Hall resistance vs. $B$ seems
to be essential to make it come out in this way. The results seem
to hint at different transport pictures for the peaks- and
valleys- of the radiation induced oscillatory magnetoresistance.

In summary, high mobility GaAs/AlGaAs devices that exhibit
vanishing $R_{xx}$ under microwave radiation also show a radiation
induced modification of the Hall effect. The modification is
characterized by (a) a small reduction in the slope of the
$R_{xy}$ vs. $B$ curve upon microwave excitation, with respect to
the dark value, (b) a reduction in the magnitude of the Hall
resistance that correlates with an increase in the diagonal
resistance $R_{xx}$. The radiation induced changes in $R_{xx}$ and
$R_{xy}$ are comparable in magnitude, although $\Delta R_{xy}$ is
small compared to $R_{xy}$, and (c) a Hall resistance correction
that disappears as the diagonal resistance vanishes. These
features seem to provide new experimental insight into this
remarkable phenomenon.\cite{3,4,5,6,7,8,9,10,12,13,15}

 \vspace{0cm}

\end{document}